# 3D E-textile for Exercise Physiology and Clinical Maternal Health Monitoring

*Junyi Zhao[1,†], Chansoo Kim[1,2,†], Weilun Li[1,†], Zichao Wen[3,†], Zhili Xiao[1], Yong Wang[1,3,*], Shantanu Chakrabartty[1,*] and Chuan Wang[1,2,*]*

[1] Department of Electrical & Systems Engineering, Washington University in St. Louis, St. Louis, Missouri 63130, United States
[2] Institute of Materials Science and Engineering, Washington University in St. Louis, St. Louis, Missouri 63130, United States
[3] Department of Obstetrics & Gynecology, Washington University in St. Louis, St. Louis, Missouri 63130, United States

† These authors contributed equally
\* Corresponding author: wangyong@wustl.edu, shantanu@wustl.edu, chuanwang@wustl.edu

**Abstract**
**Electronic textiles (E-textiles) offer great wearing comfort and unobtrusiveness, thus holding potential for next-generation health monitoring wearables. However, the practical implementation is hampered by challenges associated with poor signal quality, substantial motion artifacts, durability for long-term usage, and non-ideal user experience. Here, we report a cost-effective E-textile system (costing as low as $1 per sensor) that features 3D microfiber-based electrodes for greatly increasing the surface area. The soft and fluffy conductive microfibers disperse freely and securely adhere to the skin, achieving a low impedance at the electrode-skin interface even in the absence of gel. A superhydrophobic fluorinated self-assembled monolayer was deposited on the E-textile surface to render it waterproof while retaining the electrical conductivity. Equipped with a custom-designed motion-artifact cancelling wireless data recording circuit, the E-textile system could be integrated into a variety of smart garments for exercise physiology and health monitoring applications. Real-time multimodal electrophysiological signal monitoring, including electrocardiogram (ECG) and electromyography (EMG), was successfully carried out during strenuous cycling and even underwater swimming activities. Furthermore, a multi-channel E-textile was developed and implemented in clinical patient studies for simultaneous real-time monitoring of maternal ECG and uterine EMG signals, incorporating spatial-temporal potential mapping capabilities. Such advancement could contribute to uterus-relevant physiological assessments and early detection of abnormal uterine contraction patterns even birth-related risks, thereby improving prenatal care and providing valuable insights for women's health.**

The rapid advancement of soft wearable electronics is propelling a revolution in the field of activity tracking, health monitoring and clinical medical devices by providing greatly enhanced user experience and patient compliance. Although sensor patches that are attached on skin[1-5] or sensors and electronics that are directly fabricated on skin[6-9] are widely recognized as effective approaches, they may lack the desired unobtrusiveness and require complicated fabrication processes, thus preventing them from being ready-to-use and comfortable for long-term wearability. Textiles have emerged as a platform for seamlessly integrating various smart functionalities, including sensing[10-16], energy harvesting[17-19], displaying[20], and wireless communication[21]. Diverse materials have been explored for constructing conductors on textiles, including metals[10,12,14,17], carbon-based materials[22,23], and conjugated polymers[24,25]. Mainstream electronic textile (E-textile) manufacturing methods involve knitting/weaving/embroidering one-dimensional (1D) fiber/yarn-shaped conductive patterns into finished textile[13,14,21], or soaking/coating/printing conductive ink/paste onto fabric surface to create 2D conductive elements[10,12,17,22,24]. Especially, screen printing stands out as a promising approach due to its fast, cost-effective, and scalable patterning capability for mass production.

Real-time monitoring of electrophysiological signals is crucial in medical diagnosis, health monitoring, and exercise physiology studies. While E-textiles have great potential for applications in those scenarios, there are also a few key challenges, including maintaining a reliable skin-electrode contact for high-quality signal recording, mitigating motions artifacts from ambulatory subjects, and the necessity to eliminate the need for conductive hydrogel for long-term usability. When compared to conventional gel-assisted planar electrodes[24,26,27], gel-free dry electrodes hold great potential in addressing issues such as skin irritation, discomfort caused by tightening/pressing operation for improved contact, and signal degradation over time caused by gel dehydration. Additionally, the practical application of textile-based electronics is hindered by issues of durability and reliability. For instance, perspiration and ambient moisture can lead to electrode erosion or short circuits, resulting in performance degradation. Polymer encapsulation or lamination have been validated as effective approaches for packaging the devices, but the use of polymer layers coated on textile compromises the natural permeability, breathability, and softness of the fabrics[12,15,22,27]. Lastly, the feasibility of adopting E-textiles in real-life applications can also be influenced by manufacturing cost and consumer expenses for usage and maintenance.

Here we report a versatile 3D E-textile system (Fig. 1) featuring unique 3D microfiber-on-textile electrodes that have numerous out-of-plane conductive microfibers to allow for easy penetration through body hair and secure adhesion to the skin, thereby contributing to exceptional conformability, enlarged contact area, greatly reduced electrode-skin impedance and increased signal-to-noise ratio (SNR) in gel-free conditions, and greatly suppressed motion artifacts even in the presence of strong motion during intense sports activities. To achieve waterproof E-textiles without compromising the mechanical properties of the textile and conductivity of the electrodes, an efficient one-step surface treatment was introduced to vaporize *1H,1H,2H,2H*-perfluorooctyltrichlorosilane (PFOTS) to form a superhydrophobic self-assembled monolayer (SAM) on the E-textile surface to enable the system to be used even in underwater conditions. A custom data recording circuit has also been designed to provide real-time signal recording, processing, and wireless communication, which in combination with the E-textile, enables on-body recording of a wide range of biopotential signals, including ECG for heart and EMG for skeleton and smooth muscles. The waterproof and perspiration-tolerant E-textiles can be made into smart garments (cycling jersey and shorts, swimsuit, etc.) to provide profound insights into athletic performance and health conditions for sport physiology applications. Moreover, clinical patient studies have also been carried out using a multi-channel

E-textile system to validate the feasibility of real-time wireless monitoring of maternal ECG and uterine contractions during labor and delivery.

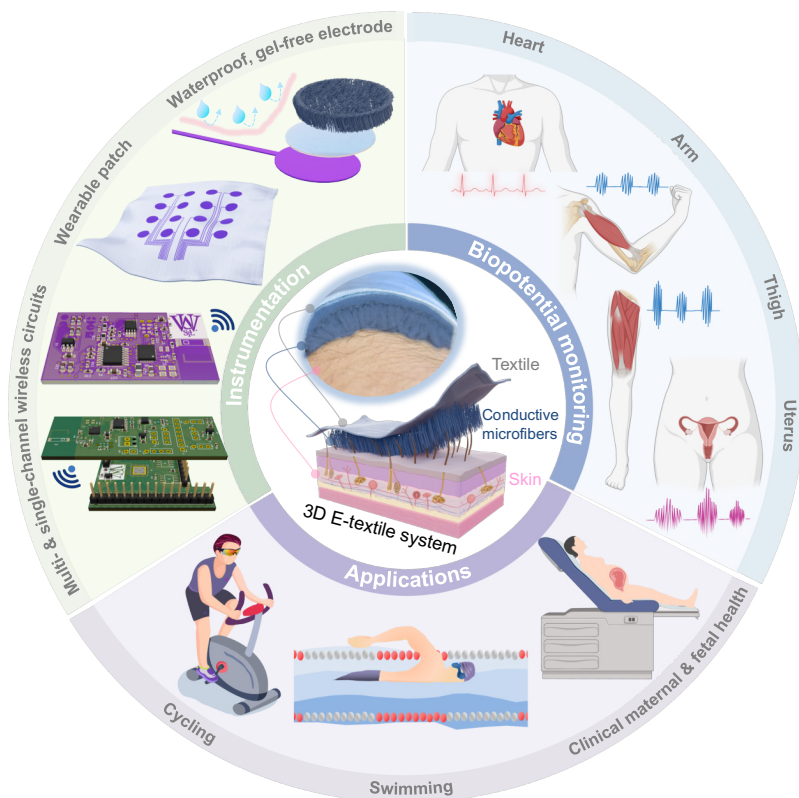

**Fig. 1 | Overview of the wireless, waterproof, and gel-free 3D E-textile system.**

## Results
### Waterproof and gel-free 3D microfiber-on-textile electrode

Conductive polymer poly(3,4-ethylenedioxythiophene) poly(styrene sulfonate) (PEDOT:PSS) offers solution processability and biocompatibility, and it has been extensively adopted as conductors for building soft and wearable electronics[28-30]. Poly(ethylene oxide) (PEO), as a high-molecular-weight and non-crosslinked polymer additive, has been shown to enhance the stretchability, conductivity, and printability of PEDOT:PSS conductors (Supplementary Fig. 1)[30-32]. Therefore, PEDOT:PSS/PEO composite ink has been formulated for screen-printed electrodes on textiles (Methods). The scanning electron microscopy (SEM) image (Fig. 2a) and energy dispersive X-ray spectroscopy (EDS) (Supplementary Fig. 2) validates the uniform coating of PEDOT:PSS/PEO composite on the textile, as compared to the pristine region. To attain good elasticity and comfortability, the nylon-spandex fabric was selected to construct the wearable E-textile. Despite the inherent biaxial stretchiness of the weft knitted yarns, the stretchability and conductive pathways of the on-textile conductors still exhibit significant variations, depending on factors such as front/back sides and wale/course directions[33]. Notably, the PEDOT:PSS/PEO conductor printed on the front side along the wale direction (Fig. 2b and Supplementary Fig. 3) exhibited superior and robust performance, with stable sheet resistance values throughout consecutive stretching tests at 5%, 15%, and 25% strain levels (Fig. 2c). To expand the applicability of the E-textile system in harsher conditions (*e.g.* user perspiration or underwater), a PFOTS self-assembled monolayer (SAM) was evenly vaporized on the outermost surface (Fig. 2d and Supplementary Fig. 4). The use of PFOTS SAM preserves the natural physical properties of yarns and the intrinsic conductivity of the on-textile PEDOT:PSS/PEO conductor (Supplementary Fig. 5), thus overcoming a substantial challenge that conventional waterproofing treatments used for textiles (polymer coating or lamination) would completely cover the conductor and render them nonconductive. Such a facile approach provides remarkable versatility for the waterproof treatment of various textiles, including knitted (cotton), woven (polyester), nonwoven (cleanroom wipes), and even cellulose-based paper (Supplementary Fig. 6). As shown in Fig. 2e, the contact angle of a water droplet on PFOTS-functionalized nylon experienced a substantial increase from 54° to 133° and the waterproof feature could last for over 7 days in ambient conditions, even after the sample was submerged completely in water for 1 hour. Furthermore, the E-textile was subjected to bending and 360° twisting underwater to showcase its excellent flexibility and durability in aqueous environment, with negligible change in sheet resistance, unlike the control sample without PFOTS treatment that exhibited a gradual degradation in the electrical property (Fig. 2f). The rapid, convenient, and eco-friendly process, requiring only micron-liter PFOTS and a minute of time, dramatically broadens the adaptability of E-textiles in a wide range of scenarios, enabling proper functioning during sweating conditions or even for water sports.

Despite the advantages of the aforementioned printed on-textile electrodes, relying solely on a planar conductive film cannot eliminate the need for conductive hydrogel in biopotential recording applications due to the high electrode-skin impedance under dry conditions (Supplementary Fig. 7). To eliminate the need for gel, conductive microfibers were developed as an add-on component on the base conductor. These out-of-plane fibrous conductors spread out and conform uniformly to the skin, effectively increasing the contact area and lowering the impedance. Considering their hairy nature, different drying strategies (Supplementary Fig. 8 and 9) and ink formulations (Supplementary Fig. 10-12) were thoroughly investigated to obtain the highly durable and reliable conductive microfibers, thereby enabling the convenient and robust biopotential measurements from dry skin. To ensure an effective bonding and electrical conduction between the on-textile conductor and the add-on conductive microfibers, multiple approaches have been studied, including stitching with PEDOT:PSS-soaked cotton threads or metal wires, and utilizing conductive binders such as silver epoxy, carbon tape, or ionic binder (Supplementary Fig. 13). The end-product of the 3D microfiber-on-textile electrode is schematically illustrated in Fig. 2g, which comprises microfibers soaked in the PEDOT:PSS/PEO composite ink (5/1 weight ratio) (Fig. 2h, i, and Supplementary Fig. 14) and dried with a hair dryer, securely attached to the on-textile conductive interconnect layout using an ionic binder. Equipped with the fluffy microfiber electrodes, the skin-electrode contact impedance under dry conditions decreased by nearly one order of magnitude and is even multiple folds lower than the commercial Ag/AgCl electrode, thereby making it possible to eliminate the need for gel (Fig. 2j). Such a significant reduction in electrode-skin impedance can be attributed to the increased contact area from microfibers, localized pressure, and enhanced adherence (Supplementary Fig. 15). All the components in Fig. 2g collectively form a versatile toolkit, where the microfiber pads can be conveniently attached or removed on demand, enabling the reusability of the conductive layout on textile.

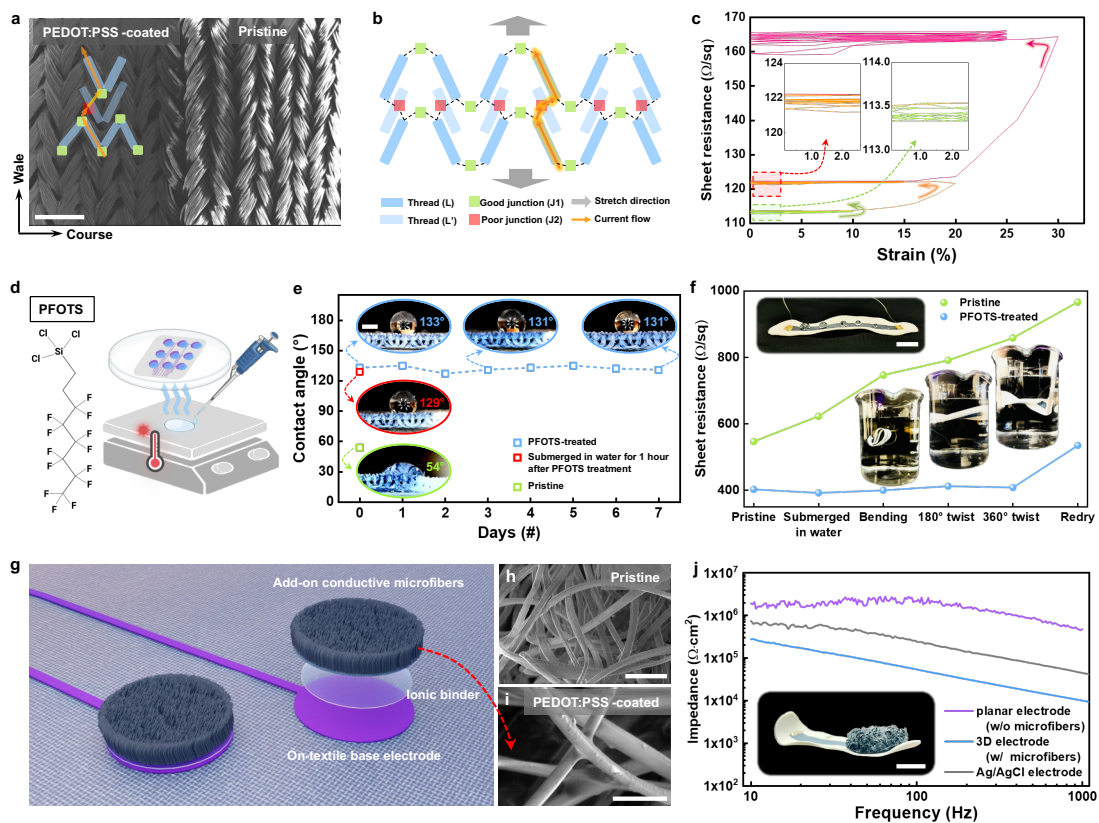

**Fig. 2 | Waterproof and gel-free 3D microfiber-on-textile electrodes. a**, Top-view SEM image of the textile interface between the PEDOT:PSS-coated (left) and non-coated (right) regions. Scale bar, 500 μm. **b**, Schematic illustrating the threads in the textile and current flow path. **c**, Evolution of sheet resistance for on-textile electrodes during multiple stretching tests of 5%, 15%, and 25% in sequence. **d**, Schematic of PFOTS-treatment to achieve waterproof E-textile. **e**, Evolution of contact angles on PFOTS-treated textile over 7 days, in comparison with the contact angle on the untreated textile and the PFOTS-treated one submerged in water for an hour. The insets show the microscopic images of water droplets on the corresponding textile. Scale bar, 500 μm. **f**, Sheet resistance of PFOTS-treated on-textile electrodes during the process of submerging in water, bending, 180° twisting, 360° twisting, and redrying. The insets show the photographs of water droplets balling up on the as-prepared PFOTS-treated textile, along with the corresponding operations under water. Scale bar, 1 cm. **g**, Schematic of the waterproof and gel-free electrode consisting of on-textile base electrode, ionic binder, and add-on conductive microfibers. **h,i**, SEM images of the non-treated microfibers (**h**) and the PEDOT:PSS-coated conductive microfibers (**i**). Scale bars, 50 μm (**h**) and 25 μm (**i**). **j**. Electrode-skin contact impedance over the frequency range of 10 to 1000 Hz under dry conditions. The inset shows the photograph of a functional single electrode made by adding conductive microfibers onto the on-textile base electrode. Scale bar, 1 cm.

**Performance of the E-textile for electrophysiological signal recording**

To verify the effectiveness of the gel-free and waterproof E-textile, we positioned a set of electrodes on forearms to capture the ECG signal under both dry and sweating conditions (Fig. 3a). The ECG amplitudes acquired by 3D microfiber-on-textile electrodes were approximately 20-fold higher in dry condition and 15-fold higher in sweating condition, as compared to the control set of solely on-textile electrodes (Fig. 3b and c). The SNR achieved by the microfiber-on-textile electrodes under dry condition is over 26.5 dB, which is substantially higher than the 8.6 dB obtained by the on-textile electrodes (Fig. 3d). Remarkably, the presence of ion-rich sweat did not diminish the signal quality; instead, it enhanced the SNR by bridging the skin and electrodes through the formation of an ionic interlayer, essentially functioning similarly to the conventional gel (Supplementary Fig. 15 and 16). For practical daily usage, the washability of the waterproof E-textile with printed conductive on-textile electrodes is crucial for its reusability as a base platform. To simulate laundry conditions, the samples were subjected to agitation in warm water for 10 min, followed by drying with a hair dryer. The SNR remained stable after 10 washing cycles, with only small fluctuations within a range of 9~12 dB (Fig. 3e and f), which validates the long-lasting waterproofing durability achieved through PFOTS treatment (Supplementary Fig. 17). Additionally, the microfiber-on-textile electrodes were subjected to testing and storage in ambient air for a duration of over 4 months, during which no noticeable degradation in SNR was observed, demonstrating the superior long-term stability and reliability (Fig. 3g and h). Furthermore, motion artifacts pose a vital challenge in the seamless integration of E-textiles with the human body, especially under gel-free conditions. Compared with the planar electrodes, the fluffy and conductive microfibers help conform to the skin robustly, functioning as a shock absorber by reducing friction and triboelectrification caused by body movement, thus effectively mitigating the motion artifacts. (Supplementary Fig. 18).

The E-textile is also capable of monitoring the electrical responses of muscles during nerve stimulation. A set of microfiber-on-textile electrodes, spaced 5 cm apart, were positioned on the arm to collect bicep-EMG signals (Fig. 3i). As the weight being lifted was increased, the EMG amplitudes increased accordingly (Fig. 3j and k) and the SNRs under dry and sweating conditions were comparable (Fig. 3l), further indicating the gel-free and waterproof features of our E-textile. Notably, our microfiber-on-textile electrodes consistently outperformed the commercial Ag/AgCl electrodes in terms of sensitivity, signal amplitude and SNR across all tests (Supplementary Fig. 19).

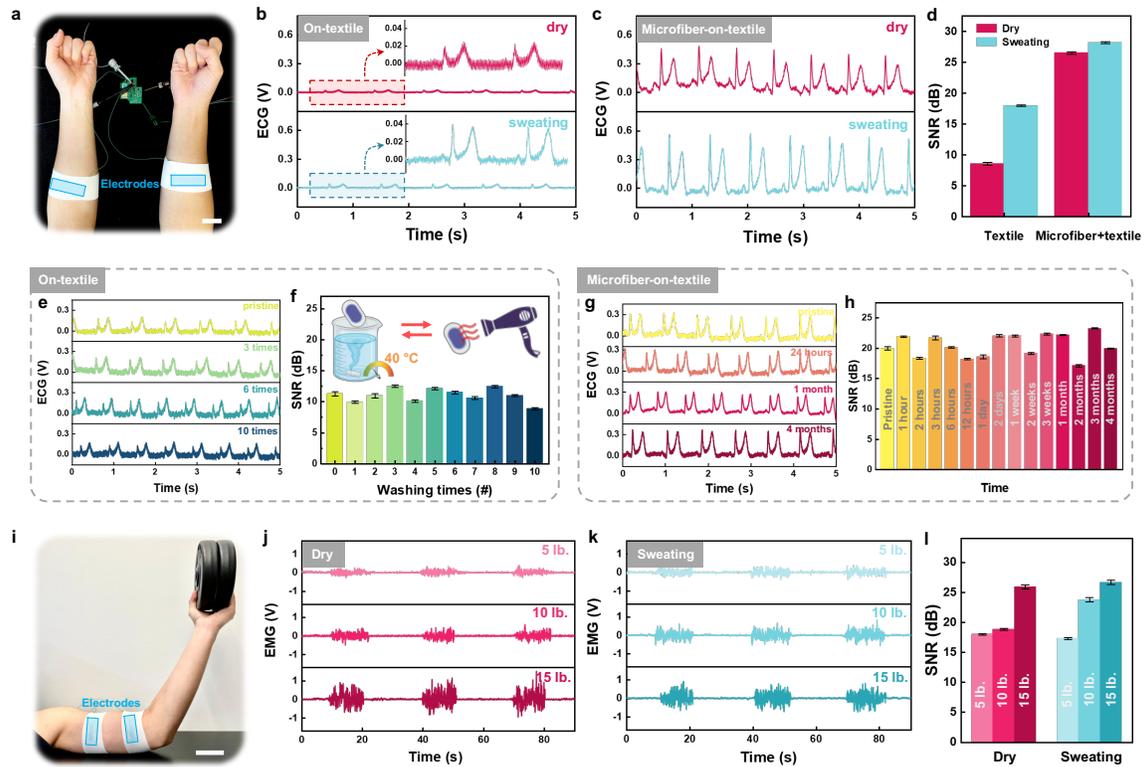

**Fig. 3 | Using E-textile for electrophysiological signal recording. a**, ECG recording by wearing the E-textile electrodes on forearms. Scale bar, 5 cm. **b,c**, ECG recording by wearing the on-textile electrodes (**b**) and the 3D microfiber-on-textile electrodes (**c**) under dry and sweating conditions. **d**, Comparison of SNR between using on-textile electrodes and microfiber-on-textile electrodes. **e,f**, Washability characterization of the on-textile electrodes by measuring ECG signals after multiple washing-redry cycles. The inset illustrates the experimental implementation of the washing test, including agitation in warm water (40 °C) and drying process. **g,h**, Long-term stability characterization of the microfiber-on-textile electrodes over a period of up to 4 months. **i**, EMG recording by wearing the electrodes on the arm. Scale bar, 5 cm. **j-l**, EMG recording by wearing the E-textile equipped with microfiber-on-textile electrodes under dry (**j**) and sweating (**k**) conditions by lifting various weights, with the corresponding SNR values compared in (**l**).

**Real-time ECG & EMG monitoring during intense cycling exercise**

Sports physiology relies on extensive real-time data collection (*e.g.* heart rate, power output, VO$_2$ Max, etc.) to guide athletes' training for enhanced performance. While real-time electrophysiological signal recording is invaluable, it presents significant challenges, particularly during intense exercise, where motion artifacts and user perspiration can pose issues. The 3D microfiber-on-textile electrodes are ideally suited for such applications. A single-channel wireless data recording system on miniaturized printed circuit board (mini-PCB, Fig. 4a) has been designed and manufactured to connect with the E-textile, thus enabling a compact and wearable system for on-demand placement, real-time signal collection and processing, and wireless communication with a user interface through a radiofrequency (RF) module (Fig. 4b). The detailed circuit diagram of the analog frontend and the cutoff frequency calculation are presented in Methods and Supplementary Fig. 20. To demonstrate the practical application of the E-textiles for exercise physiology, a cyclist wore the cycling jersey equipped with microfiber-on-textile electrodes (symmetrically positioned on the inner side facing the lower chest, Fig. 4c) while cycling on a stationary bike (Fig. 4d). Continuous and real-time monitoring of ECG signals was conducted throughout a 14-minute training session, comprising a 4.5-minute warm-up phase, a 5-minute climbing phase, and a 4.5-minute sprinting phase (Fig. 4f(i) and Supplementary Video 1). Meanwhile, heart rate (HR) analysis was performed to monitor the real-time HR for zone-based training (Fig. 4f(v)). During the initial stage, the subject began in a relaxed state of pedaling for warming up and the recorded ECG signals exhibited exceptional stability (Fig. 4f(ii)), where the HR averaged at around 120 beats per minute (bpm). As the climbing phase began, the subject exerted greater efforts and lowered the upper body to overcome the increased pedaling resistance, resulting in a gradual rise in HR to 160 bpm (Fig. 4f(iii)). The motion artifacts remained negligible in this phase with only occasional voltage spikes that had no discernible influence on the signal quality. During the sprinting phase, the subject had to move his body left-and-right vigorously and the presence of motion artifacts became more noticeable. Nevertheless, even under such extreme condition, the unprocessed raw ECG signal still preserves all distinct ECG features with clear QRS complex and T-wave (Fig. 4f(iv)). Moreover, despite the subject's intense perspiration throughout the climbing and sprinting phases (Fig. 4e), the waterproof durability of the E-textile enabled uninterrupted recording.

EMG is another crucial electrophysiological signal that athletes monitor to assess muscle output and track training performance. During the experiment, the cyclist wore a E-textile-based cycling short with microfiber-on-textile electrodes positioned 5 cm apart facing the thigh to monitor the electrical activities of the quadriceps muscle (Fig. 4g). The thigh-EMG signals were monitored in real-time during a 6-minute session, comprising four phases: warm-up, climbing, sprinting, and cool-down. It is worth noting that performing on-body real-time EMG recording, particularly in gel-free conditions, poses extreme challenges due to intense frictions and deformations between the skin and the electrodes caused by vigorous body movements and localized muscle contractions (Supplementary Video 2). The raw thigh-EMG signals collected by our E-textile (Fig. 4i(i)) and the zoomed-in views (Fig. 4(ii)-(v)) clearly indicated the evolution of muscle contraction activities during different exercise phases. Meanwhile, additional data analysis was conducted to extract valuable insights, allowing the cyclist to make timely and professional assessments of exercise performance and muscle activity. Cadence is a vital metric in cycling training used to track the pedaling rate to guide the cyclists in maintaining effective power output and minimizing the risk of overuse injury. To derive cadence information from the raw EMG signal, the following steps were employed: applying the bandpass filtering (1~50 Hz) to isolate the target frequency range, utilizing the maximal overlap discrete wavelet transform (MODWT) algorithm to do segmentation, identifying the peaks corresponding to muscle contractions, and calculating the interval between consecutive peaks as the duration per pedal revolution. The cadence extracted by the E-textile precisely aligned with the data recorded by the commercial sensing module mounted on the stationary bike's crank arm (Fig. 4i(vi)(vii)). Besides cadence, the EMG amplitude analysis provides information about the muscle activation and exertion during the training. The extracted EMG amplitude exhibited a proportional correlation with the digital resistance setting on the bike (Fig. 4i(viii)(ix)). During the warm-up phase of low-resistance pedaling, the EMG amplitude remains fairly low and stable with clear plateau patterns that match the bike's resistance setting, indicating sustained and consistent muscle output for generating a constant force. When dialing up the resistance to enter the climbing phase, the subtle delay followed by an abrupt rise in the EMG amplitude suggests the presence of hysteresis in muscle activation and neuromuscular coordination. When the subject was struggling in the sprinting phase, the fluctuation and instability in the EMG amplitude could be attributed to intensified muscle contraction and vigorous body swinging required to overcome the over-loaded resistance. This unstable pattern persists until reaching volitional fatigue, followed by the transition into the cool-down phase with a drop in EMG amplitudes and subsequent stability.

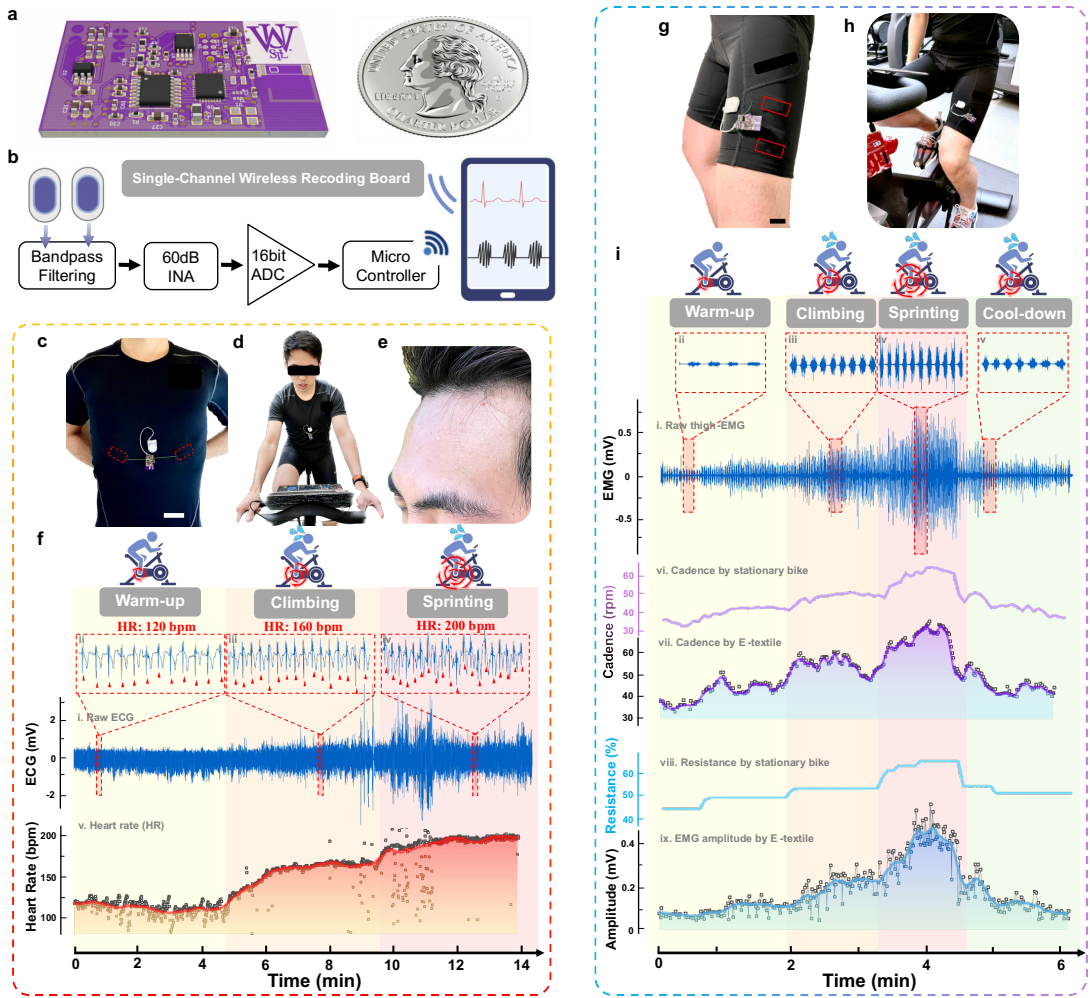

**Fig. 4 | E-textile for real-time ECG & EMG monitoring and data analysis during cycling training. a**, Single-channel mini-PCB for the recording of electrophysiological signals during intensive exercise. **b**, Block diagram of the portable instrumentation system for recording electrophysiological signals. **c,d**, Photographs of the subject wearing an E-textile-based cycling jersey for cycling on a stationary bike. The red-circled regions indicate the placement locations of electrodes. Scale bar, 5 cm. **e**, Photograph showing the subject experienced substantial sweating during the exercise. **f**, Real-time ECG monitoring and heart rate analysis during a 14-minute cycling session including the warm-up, climbing, and sprinting phases. Recorded raw ECG signals (i). The zoomed-in views of the ECG signals within 6-second windows during different training phases (ii-iv). The real-time heart rate analysis throughout the training session (v). **g,h**, Photographs of the subject wearing E-textile-based cycling shorts. The red-circled regions indicate the placement locations of electrodes. Scale bar, 2 cm. **i**, Real-time EMG monitoring and muscle output analysis of the subject's quadricep during a 6-minute cycling session, including the warm-up, climbing, sprinting, and cool-down phases. Recorded raw thigh-EMG signals (i). The zoomed-in views of the EMG signals within 10-second windows during different training phases (ii-v). Training information from the stationary bike including cadence (vi) and resistance (viii), and real-time muscle output analysis from the E-textile system including cadence (vii) and EMG amplitude (ix).

**Real-time ECG monitoring during swimming exercise**

Despite commercial products such as swim watches, chest straps, and arm bands have emerged for monitoring heart rate (HR) during water sports, they are limited to recording HR at discrete time intervals. Achieving on-body real-time and continuous recording of raw electrophysiological signals in an underwater environment remains an unsolved challenge. The substantial interference caused by water renders the use of electrodes underwater impractical, further complicating the accurate biopotential recording. By integrating our waterproof and gel-free E-textile system on a swimsuit (Fig. 5a), real-time and continuous monitoring of the swimmer's ECG became feasible. We have conducted a 11-minute recording session comprising various activities, including walking from the locker room to the pool, stepping into the pool, floating and diving, swimming, exiting the pool, and walking back to the locker room (Fig. 5b). During the initial 3.5 minutes, after putting on the E-textile-equipped swimsuit and a brief shower to moisturize the skin, the swimmer proceeded to walk towards the pool, during which a nearly motion-artifact-free ECG recording could be observed (Fig. 5b(ii)). When stepping into the pool, significant motion was introduced during the process, which coupled with water gradually permeating through the swimsuit, causes observable interference with the sensing electrodes facing the swimmer's back. The interaction with water and friction at the electrode-skin interface resulted in motion artifacts and noticeable but small voltage spikes. Nevertheless, upon closer examination of the raw signals captured, the characteristic ECG features such as the QRS complex remained distinct even during the noisiest phase of the process (Fig. 5b(iii)). During floating and diving practices to acclimate to the water environment and warm up, the sensing module recorded consistently without disruption even when the swimmer was completed submerged underwater (Fig. 5b(iv)). Afterwards, the participant performed normal swimming activities using breaststroke style along the lane for several rounds, and the recorded signal quality remained unaffected by the intense body motions and water interference (Fig. 5b(v)). As the swimmer exited the pool, Minor motion artifacts could once again be observed in the recording, but the ECG signals immediately recovered and stabilized during the walk back to the locker room (Fig. 5b(vi)(vii)). Overall, the successful implementation of E-textile for underwater electrophysiological signal recording showcases a remarkable advancement in technology for water sports physiology, providing a promising tool for elevating training effectiveness.

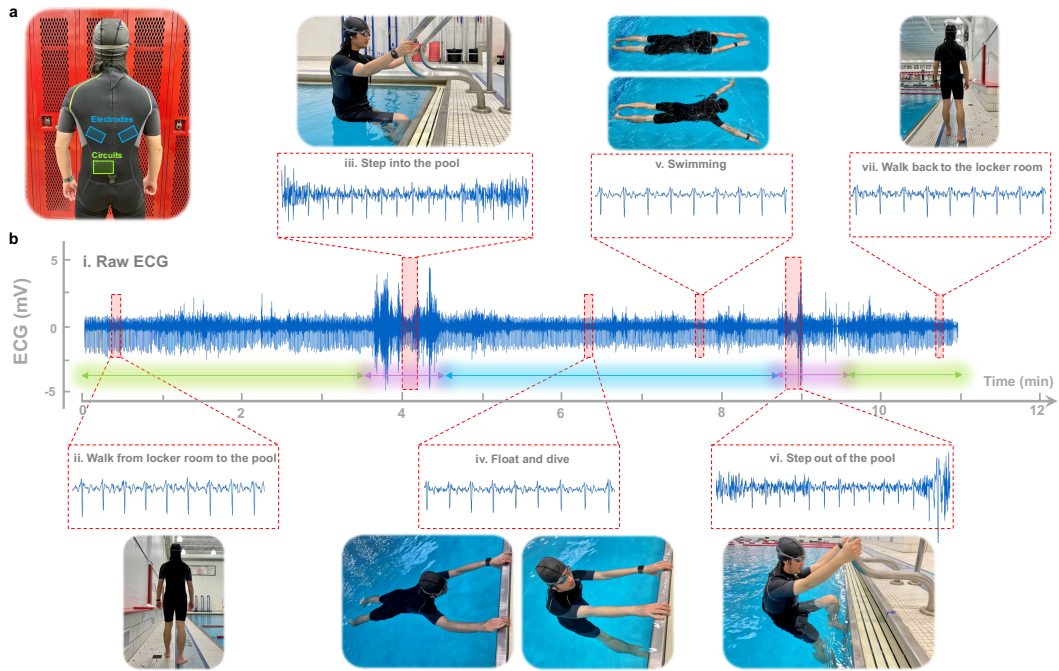

**Fig. 5 | E-textile for real-time ECG monitoring during the swimming training. a**, Photograph of the subject wearing the E-textile-based swimsuit in the locker room. The blue- and green-circled regions indicate the placement locations of the electrodes and portable mini-PCB, respectively. **b**, Real-time ECG monitoring of a swimmer throughout an entire 11-minute swimming training session, starting from walking to the pool from the locker room (green range on the left), stepping into the pool (purple range on the left), floating and diving and swimming (blue range), stepping out of the pool (purple range on the right), and walking back to the locker room (green range on the right). Raw ECG recording (i) and zoomed-in views of the corresponding stages (ii-vii).

**Real-time clinical maternal health monitoring during labor**

In obstetrics, uterine contractions are usually recorded clinically using pressure-sensitive tocodynamometer (TOCO) or intrauterine pressure catheter (IUPC). However, these "gold-standard" instruments sometimes have limited signal accuracies due to incorrect instrument placement and causes significant patient discomfort due to the tight bandages used for TOCO or the invasiveness from IUPC. The uterine EMG signal plays an essential role in assessing and interpreting the uterine contraction activities, providing valuable aid in diagnosing labor dysfunction and predicting preterm labor. However, real-time acquisition of uterine EMG signals can be challenging due to the dynamic nature of uterine contractions and the complexity of achieving acceptable SNR due to the low EMG amplitude produced by smooth muscles, the deep-lying location of the uterus, the interference by electrophysiological signals from neighboring organs and heart, and the maternal body movement. To use the E-textile system in uterine contraction monitoring, the single-channel recording module has been upgraded to a multi-channel version with additional multiplexing functionality at the frontend (Fig. 6a). This instrumentation integrates a separate daughter-PCB beneath the mother-PCB to incorporate a differential analog multiplexer and active shielding circuits, thereby enabling simultaneous recording from up to 16 channels (Supplementary Fig. 21). As illustrated in Fig. 6b, an E-textile patch with a printed electrode array layout, connected to the stacked PCBs, was attached to the maternal abdomen during the experiment. This configuration also offers wireless transmission of signals to a receiving device located outside the labor & delivery room, enabling real-time multi-channel signal recording, monitoring, and analysis. During the clinical patient study, multiple instruments, including an 8-channel E-textile system, the TOCO system, and a commercial biopotential measurement system with active Ag/AgCl electrodes (BioSemi), were implemented together on pregnant subjects to cross-validate the effectiveness and quality of recordings (Methods and Fig. 6c,d). In addition to uterine EMG, maternal ECG could also be captured simultaneously from the same electrodes, revealing the coupling of multiple biopotential signals originating from various organs and tissues embedded in the raw recording. Therefore, a dedicated protocol for data interpretation has been developed to enable multimodal monitoring of maternal health. Take the recording conducted on Subject #1 as an example (Extended Data Fig. 1), the wirelessly received signals were initially demultiplexed into separate channels, followed by median filtering, crosstalk compensation, bandpass filtering, and motion-artifact cancelation to enhance the signal quality (Fig. 6a). Fast Fourier Transform (FFT) analysis was then performed to extract feature components in frequency domain, ultimately contributing to the identification of dominant frequency peaks and distributions associated with maternal ECG (1~3 Hz) and uterine EMG (0.3~1 Hz). Following this protocol, the real-time ECG recorded throughout the entire 21-minute clinical session is presented in Fig. 6e(i), revealing high-quality signals with clear ECG features in the zoomed-in windows. The real-time HR analysis, averaging at 87 bpm, reflects regular cardiac cycles and rhythm of the subject and aligns well with the dominant frequency (1.45~1.5 Hz) observed in the time-frequency analysis (Fig. 6e(ii)(iii)).

Regarding the uterine EMG monitoring, eight uterine contractions were confirmed using the TOCO, along with some irregular spikes likely caused by maternal movements (Fig. 6f(iv)). The raw EMGs recorded by the E-textile and BioSemi closely resemble each other in terms of amplitude and SNR (Fig. 6f(i)(iii)). Moreover, the root-mean-square (RMS) waveform derived from the E-textile signal (Fig. 5f(ii)) correlate excellently with the TOCO waveform. To validate the reproducibility of E-textile in clinical settings, it was tested in a total of four clinical patient studies by involving multiple pregnant subjects (Subjects #1-4). The E-textile successfully accomplished all patient studies, yielding high-quality multimodal recordings. In contrast, the commercial BioSemi using rigid electrodes missed some feature contractions for Subject #2 (Extended Data Fig. 2) and failed to record due to the patch delamination caused by Subject #3's vigorous movements (Extended Data Fig. 3). The TOCO also failed to record due to the displacement induced by Subject #4's motion (Extended Data Fig. 4). Notably, these four patients represented a diverse range of characteristics, including variations in age (19~28 years old), delivery history (nulliparous or multiparous), labor type (induction or spontaneous), and body conditions (body mass index), providing compelling evidence to evaluate the effectiveness and robustness of the E-textile (Extended Data Table 1,2).

Moreover, the E-textile's ready-to-use and disposable attribute eliminate tedious workload associated with the electrode preparation and post-cleaning needed for the BioSemi in clinical studies (Supplementary Fig. 23). This simplifies the overall usage procedure and could potentially allow users to apply the E-textile themselves for in-home monitoring applications. More significantly, when compared to the state-of-the-art uterine contraction monitoring equipment that costs over tens of thousands of US dollars, our E-textile system stands out as a highly affordable alternative. The reusable wireless recording hardware costs around $93, while the disposable patch costs less than $5 (Supplementary Table 1 and 2). The affordability of this low-cost E-textile system paves the way for widespread adoption of the technology for both clinical and in-home maternal health monitoring and labor management practices, especially in low-resource settings.

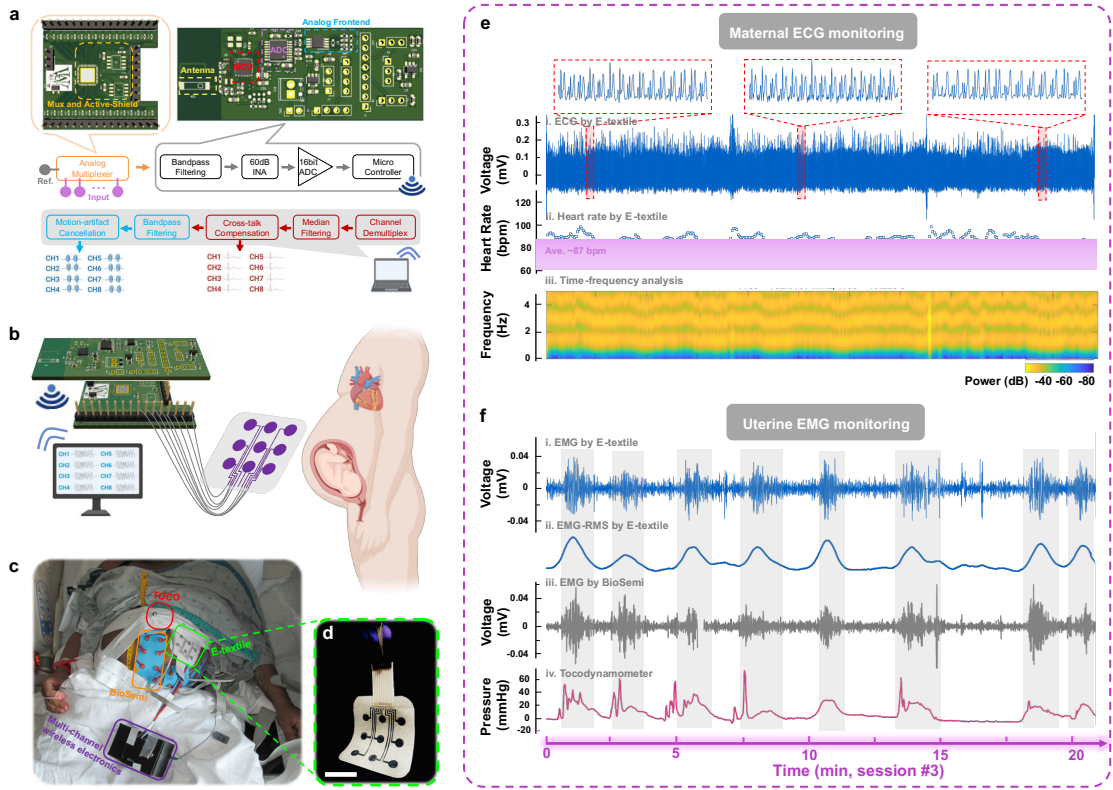

**Fig. 6 | E-textile for real-time maternal health monitoring in a clinical setting. a**, System-level block diagrams of the multi-channel E-textile system, including the multi-channel signal transduction (orange), signal processing and conditioning (grey), wireless data transmission to the portable electronic device, channel demultiplexing, and corresponding post-processing to display the readable multi-channel ECG (red) and multi-channel EMG (blue) signals. The insets show the daughter board (top left) for active shielding and analog multiplexing; and the mother board (top right) with analog frontend for signal amplification, analog-to-digital conversion (ADC), and antenna for wireless transmission. **b**, Schematic illustration depicting the E-textile patch equipped with the multi-channel data recording electronics, designed for real-time maternal health monitoring in clinical settings by attaching the system to the subject's abdomen. **c**, Photograph showing multiple instrumentations being used in a clinical patient study to monitor the uterine contraction activities during the laboring process, including TOCO (red), BioSemi (orange), and E-textile (green) equipped with multi-channel wireless recording electronics (purple). **d**, Photograph of the E-textile patch with an array of 9 electrodes used in the clinical study. Scale bar, 5 cm. **e**, Real-time maternal ECG monitoring and data analysis (*e.g.* real-time heart rate and time-frequency analysis) of the pregnant subject. Raw maternal ECG signals (i), the extracted heart rate (ii), and the time-frequency analysis (iii). **f**, Real-time uterine EMG monitoring (i) and RMS analysis (ii) from the wireless E-textile system, in comparison with the recordings from commercial BioSemi system (iii) and the tocodynamometer (iv).

## Discussion

We have demonstrated an E-textile system that integrates microfiber-on-textile electrodes and miniaturized circuitry, enabling comfortable and wearable electrophysiology monitoring. The free-standing microfibers allow for on-body recordings of multiple biopotential signals in gel-free and motion-artifact-tolerant manners. The water-repellence of self-assembled PFOTS monolayer equipped the E-textile sweat-tolerant and waterproof capability, ensuring uninterrupted electrophysiology monitoring even in conditions of severe perspiration or underwater usage. The E-textile system, consisting of planar printed on-textile interconnects, add-on conductive microfibers, ionic binder, and mini-PCBs, could be easily assembled and integrated into garments of various form factors. This versatility allows for diverse sports physiology applications in real-life scenarios such as strenuous cycling and water sports. Besides, clinical studies have validated the E-textile system's capability for multimodal monitoring of maternal health, including maternal ECG, uterine EMG, and potential distribution maps.

For future studies, the spatial coverage and temporal resolution of multi-channel E-textile could be further improved by performing simultaneous recording from multimodal sensors (physical and electrophysiological sensing modalities) and increasing the sampling rate to derive a comprehensive and detailed physiological assessment, serving for applications in biopotential mapping[5,34]. Meanwhile, customization of electrode size, distribution, and layout may be tailored to suit the specific needs of individual users and various body parts, thereby facilitating personalized healthcare solutions. Furthermore, the transformation of the E-textile into affordable and unobtrusive wearable instruments holds potentials for professionally assessing sports performance, monitoring of maternal/fetal health, and facilitating data-driven medical diagnosis.

**References**
1    Gao, W. *et al.* Fully integrated wearable sensor arrays for multiplexed in situ perspiration analysis. *Nature* **529**, 509-514 (2016).
2    Wang, M. *et al.* A wearable electrochemical biosensor for the monitoring of metabolites and nutrients. *Nat Biomed Eng* **6**, 1225-1235 (2022).
3    Bariya, M., Nyein, H. Y. Y. & Javey, A. Wearable sweat sensors. *Nature Electronics* **1**, 160-171 (2018).
4    Yang, Y. *et al.* A laser-engraved wearable sensor for sensitive detection of uric acid and tyrosine in sweat. *Nat Biotechnol* **38**, 217-224 (2020).
5    Li, W., Zhao, J., Wang, Y., Wang, C. & Chakrabartty, S. A Low-Power Impedance-to-Frequency Converter for Frequency-Multiplexed Wearable Sensors. *IEEE Trans Biomed Circuits Syst* **PP** (2024).
6    Ershad, F. *et al.* Ultra-conformal drawn-on-skin electronics for multifunctional motion artifact-free sensing and point-of-care treatment. *Nat Commun* **11**, 3823 (2020).
7    Patel, S. *et al.* Drawn-on-Skin Sensors from Fully Biocompatible Inks toward High-Quality Electrophysiology. *Small* **18**, e2107099 (2022).
8    Kim, K. K. *et al.* A substrate-less nanomesh receptor with meta-learning for rapid hand task recognition. *Nature Electronics* (2022).
9    Ershad, F., Patel, S. & Yu, C. Wearable bioelectronics fabricated in situ on skins. *npj Flexible Electronics* **7** (2023).
10   Ma, Z. *et al.* Permeable superelastic liquid-metal fibre mat enables biocompatible and monolithic stretchable electronics. *Nat Mater* **20**, 859-868 (2021).
11   Chen, G., Fang, Y., Zhao, X., Tat, T. & Chen, J. Textiles for learning tactile interactions. *Nature Electronics* **4**, 175-176 (2021).
12   Matsuhisa, N. *et al.* Printable elastic conductors with a high conductivity for electronic textile applications. *Nat Commun* **6**, 7461 (2015).

13  Leber, A. *et al.* Soft and stretchable liquid metal transmission lines as distributed probes of multimodal deformations. *Nature Electronics* **3**, 316-326 (2020).
14  Luo, Y. *et al.* Learning human–environment interactions using conformal tactile textiles. *Nature Electronics* **4**, 193-201 (2021).
15  Chen, X. *et al.* Fabric-substrated capacitive biopotential sensors enhanced by dielectric nanoparticles. *Nano Research* **14**, 3248-3252 (2021).
16  Zhao, J. *et al.* TouchpadAnyWear: Textile-Integrated Tactile Sensors for Multimodal High Spatial-Resolution Touch Inputs with Motion Artifacts Tolerance. *Proceedings of the 37th Annual ACM Symposium on User Interface Software and Technology*, 1-14 (2024).
17  Yin, L. *et al.* A self-sustainable wearable multi-modular E-textile bioenergy microgrid system. *Nat Commun* **12**, 1542 (2021).
18  Zhao, Z. *et al.* Machine-Washable Textile Triboelectric Nanogenerators for Effective Human Respiratory Monitoring through Loom Weaving of Metallic Yarns. *Adv Mater* **28**, 10267-10274 (2016).
19  Zhang, Y. *et al.* Destructive-Treatment-Free Rapid Polymer-Assisted Metal Deposition for Versatile Electronic Textiles. *ACS Appl Mater Interfaces* **14**, 56193-56202 (2022).
20  Shi, X. *et al.* Large-area display textiles integrated with functional systems. *Nature* **591**, 240-245 (2021).
21  Lin, R. *et al.* Digitally-embroidered liquid metal electronic textiles for wearable wireless systems. *Nat Commun* **13**, 2190 (2022).
22  Huang, C.-Y. & Chiu, C.-W. Facile Fabrication of a Stretchable and Flexible Nanofiber Carbon Film-Sensing Electrode by Electrospinning and Its Application in Smart Clothing for ECG and EMG Monitoring. *ACS Applied Electronic Materials* **3**, 676-686 (2021).
23  Liu, L., Yu, Y., Yan, C., Li, K. & Zheng, Z. Wearable energy-dense and power-dense supercapacitor yarns enabled by scalable graphene-metallic textile composite electrodes. *Nat Commun* **6**, 7260 (2015).
24  Takamatsu, S. *et al.* Direct patterning of organic conductors on knitted textiles for long-term electrocardiography. *Sci Rep* **5**, 15003 (2015).
25  Clevenger, M., Kim, H., Song, H. W., No, K. & Lee, S. Binder-free printed PEDOT wearable sensors on everyday fabrics using oxidative chemical vapor deposition. *Science Advances* **7**, eabj8958 (2021).
26  Bihar, E. *et al.* Fully Printed Electrodes on Stretchable Textiles for Long-Term Electrophysiology. *Advanced Materials Technologies* **2** (2017).
27  Homayounfar, S. Z. *et al.* Multimodal Smart Eyewear for Longitudinal Eye Movement Tracking. *Matter* **3**, 1275-1293 (2020).
28  Lo, L. W. *et al.* A Soft Sponge Sensor for Multimodal Sensing and Distinguishing of Pressure, Strain, and Temperature. *ACS Appl Mater Interfaces* **14**, 9570-9578 (2022).
29  Lo, L. W. *et al.* Stretchable Sponge Electrodes for Long-Term and Motion-Artifact-Tolerant Recording of High-Quality Electrophysiologic Signals. *ACS Nano* **16**, 11792-11801 (2022).
30  Zhao, J. L., L. W.; Yu, Z.; Wang, C. Handwriting of perovskite optoelectronic devices on diverse substrates. *Nature Photonics* **17** (2023).
31  Lo, L. W. *et al.* An Inkjet-Printed PEDOT:PSS-Based Stretchable Conductor for Wearable Health Monitoring Device Applications. *ACS Appl Mater Interfaces* **13**, 21693-21702 (2021).

32  Zhao, J. *et al.* High-Speed Fabrication of All-Inkjet-Printed Organometallic Halide Perovskite Light-Emitting Diodes on Elastic Substrates. *Adv Mater* **33**, 2102095 (2021).
33  Wu, L. *et al.* The deformation behaviors and mechanism of weft knitted fabric based on micro-scale virtual fiber model. *International Journal of Mechanical Sciences* **187** (2020).
34  Li, W. *et al.* A Portable and a Scalable Multi-Channel Wireless Recording System for Wearable Electromyometrial Imaging. *IEEE Trans Biomed Circuits Syst* **17**, 916-927 (2023).

32  Zhao, J. *et al.* High-Speed Fabrication of All-Inkjet-Printed Organometallic Halide Perovskite Light-Emitting Diodes on Elastic Substrates. *Adv Mater* **33**, 2102095 (2021).
33  Wu, L. *et al.* The deformation behaviors and mechanism of weft knitted fabric based on micro-scale virtual fiber model. *International Journal of Mechanical Sciences* **187** (2020).
34  Li, W. *et al.* A Portable and a Scalable Multi-Channel Wireless Recording System for Wearable Electromyometrial Imaging. *IEEE Trans Biomed Circuits Syst* **17**, 916-927 (2023).

**Acknowledgements**

This work was funded by the Bill & Melinda Gates Foundation under award numbers INV-005417 and INV-035476 and NIH/Eunice Kennedy Shriver National Institute of Child Health and Human Development under award number R01HD105905. The authors also acknowledge the Institute of Materials Science and Engineering at Washington University for the use of instruments and staff assistance.

**Author contributions**

J.Z. and C.W. conceived the idea and designed the overall studies. J.Z. and C.K. performed all the in-lab experiments, including the ink development, material characterization, device fabrication, device characterization, and biopotential recordings. W.L., Z.X., and S.C. designed and fabricated the circuits. J.Z., C.K., and W.L. performed the exercise physiology experiments. J.Z., Z.W., and W.L. conducted the clinical patient studies. J.Z., C.K., W.L., Z.W., S.C., Y.W., and C.W. conducted the data analysis. J.Z. and C.W. led the figure design and paper writing while all authors provided feedback.